\documentclass[runningheads,orivec]{llncs}

\usepackage[T1]{fontenc}
\usepackage{cite}
\usepackage{amsmath,amssymb,amsfonts}
\usepackage{algorithmic}
\usepackage{graphicx}
\usepackage{textcomp}
\usepackage{xcolor}
\usepackage{booktabs}
\usepackage{wrapfig}
\usepackage{floatrow}

\usepackage[type={CC},modifier={by},version={4.0},imagemodifier={-80x15},{hyperxmp=false}]{doclicense}

\usepackage[hyphens]{url}
\usepackage[hidelinks]{hyperref}

\usepackage{orcidlink}
\usepackage[misc,geometry]{ifsym}

\usepackage{makecell}
\usepackage{tabularx}
\usepackage{multicol}
\usepackage{enumitem}

\usepackage{color}

\newcommand{\yes}{\checkmark}
\newcommand{\no}{\boldmath $\times$}
\newcommand{\na}{$\cdot$}
\newcommand{\fail}{\boldmath $\otimes$}
\newcommand*\rot{\rotatebox{90}}

\def\xsect#1{\vspace{0.4em}\noindent\textbf{#1:}}

\begin{document}

\title{Linux disk encryption and self-encrypting drives}
\subtitle{A case study on Opal2 drives security}

\author{Milan Brož\inst{1,2}\,\Letter\,\orcidlink{0000-0001-7179-0386} \and
Tamara Čierniková\inst{1} \and Ondřej Kozina\inst{3} \and \\Vladimír Sedláček\inst{1}\,\orcidlink{0000-0003-2409-661X}}
\authorrunning{M. Brož, T. Čierniková, O.Kozina and V. Sedláček}

\institute{
Masaryk University, Brno, Czechia\\\email{\{milan.broz,ciernikova,vlada.sedlacek\}@mail.muni.cz}
\and
OpenSSL Corporation
\and
Red Hat Czech\\\email{okozina@redhat.com}
}
\maketitle

\begin{center}
\doclicenseText
\end{center}

\begin{abstract}
Opal2 self-encrypting drives provide hardware-based disk encryption serving as an additional layer of protection, or a~replacement, for software-based solutions. This paper presents a case study of real-world Linux integration of Opal2 drives and the security of Opal2 firmware. The study was conducted on a testbed of 38 commercial off-the-shelf Opal2 drives from various vendors using a black-box approach. We identified several firmware security issues and incompatibilities, which we responsibly disclosed to respective vendors. Our findings led to improvements in Linux disk encryption tools used across all major Linux distributions. To enable independent evaluation for the public, we release our test scenarios for Opal2 drives as an open-source toolset.
\end{abstract}

\keywords{SED Opal2, LUKS2, self-encrypting drives, disk encryption}

\section{Introduction}
Disk encryption is a common mechanism for protecting data at rest.
It can be implemented in software, where the operating system handles encryption per disk sector, or in hardware through \emph{self-encrypting drives} (SEDs).
The TCG Opal2 standard~\cite{tcgopal} defines a widely adopted specification for SEDs that offloads cryptographic operations to the drive controller itself.

 Opal2 can be deployed alongside software encryption as an additional security layer, or as a standalone solution where data confidentiality relies entirely on the hardware. Both use cases expect the underlying hardware to behave correctly and as specified; an assumption we put to the test.

 Linux disk encryption relies on the open-source Linux Unified Key Setup 2 (LUKS2) format~\cite{luks2} together with the \emph{cryptsetup}~\cite{cryptsetup} user-space configuration utility. The practical contribution of this work is the secure integration of Opal2 support into LUKS2, thereby enabling a user-friendly configuration of SEDs on Linux, which was not feasible with prior tooling.

We present a case study on the security of Opal2 drives. To this end, we assemble a testbed of 38 drives advertising Opal2 support and conduct a security analysis of intercepted firmware. Our testbed is a mix of second-hand units and new in-stock purchases, representing the
diversity of Opal2 SEDs found in present Linux deployments. We employ a \emph{black-box} analysis methodology: we rely exclusively on documented protocols, without any knowledge of the internal firmware implementation.

Our security evaluation focuses on non-adversarial operational use of SEDs, deliberately setting aside hardware-level or supply-chain attacks. The \emph{black-box} approach has a key advantage: every test is fully reproducible on any drive that exposes the required Opal2 interface. Importantly, our findings are not Linux-specific; they apply to any platform that supports Opal2 drives.

Analysis uncovered firmware-level issues affecting the correctness of encryption, including a discrepancy between the encryption block size and the physical sector size, potential reuse of sector tweak values, and insufficient randomness in the exported random number generator interface.
All issues were reported to affected vendors through a responsible disclosure process.

We make following contributions:
\begin{itemize}
\item \emph{A black-box security analysis of 38 Opal2 drives, revealing firmware vulnerabilities, specification ambiguities, and implementation incompatibilities. The Opal2 tooling developed for this analysis is released as open source.}
\item \emph{Backed by this analysis, we achieve Opal2 integration in LUKS2, yielding concrete security improvements to Linux's open-source disk encryption stack.}
\end{itemize}

The paper is structured as follows. Section~\ref{sec:fde} establishes disk encryption terminology, and Section~\ref{sec:seds} covers the relevant standards, features, and hardware limitations of SEDs. Section~\ref{sec:luks2} introduces our LUKS2 use case and simplified security goals. Section~\ref{sec:testing} describes our methodology, testbed, tooling, and the results of the black-box analysis of Opal2 SEDs. Practical impact is discussed in Section~\ref{sec:practical}. A formal description of our pattern analysis algorithm is provided in Appendix~\ref{sec:patterndesc}, and the full list of tested drive models and firmware versions appears in Appendix~\ref{sec:drives}.

\section{Sector-based (disk) encryption}\label{sec:fde}

Disk encryption operates on sectors provided by the drive. A disk sector is an independently addressable atomic storage unit, typically 512 or 4096 bytes in size. Sector contents are encrypted on write and decrypted on read using a symmetric cipher with a suitable encryption mode, such as AES-XTS~\cite{IEEE1619-2025,nistxts}.

Such storage encryption is referred to as Full Disk Encryption (FDE) when the entire device is encrypted, or as Volume Encryption when only a portion of the device is encrypted. In both cases, the encrypted area is a contiguous, linearly addressable storage region, such as a disk partition or a Locking Range (LR) in self-encrypting drives. The encrypted area uses a single encryption key (or a fixed set of keys), commonly referred to as a Media Encryption Key (MEK), Disk Encryption Key (DEK), or Volume Key (VK).

Disk encryption can be implemented as a driver at the operating system level, optionally leveraging hardware accelerators. Typical examples of software disk encryption include Microsoft BitLocker~\cite{bitlocker}, the VeraCrypt~\cite{veracrypt}, and the Linux-based LUKS~\cite{luks2}.

SEDs move the encryption into the drive hardware itself. A prominent example of an SED standard is the Trusted Computing Group (TCG) Opal2 specification~\cite{tcgopal}, implemented by various hardware vendors.

\section{Self-encrypting drives}\label{sec:seds}

The practical advantage of SED is that the drive can achieve native storage hardware performance with minimal power consumption, as the encryption chip in the drive can be highly optimized for the specific scenario.

To better understand the context of SED usage, we describe the limiting factors associated with commercial off-the-shelf hardware. Linux, in general, does not compel users to upgrade hardware frequently, giving them greater flexibility to use existing hardware to its full potential. Users can run second-hand hardware (such as laptops) with data protected by disk encryption. With the recent surge in AI-driven storage demands and rising prices, the use of older storage hardware is unlikely to decline quickly. This is why we also focus on hardware that is no longer supported by vendors but is still used in production systems for storing sensitive data.

Disk encryption can be applied to both internal drives (operating system) and external drives (portable disks). Hardware limitations can affect not only performance but also security.

\subsection{Interfaces and form factors}\label{sec:ff}
We consider only solid-state drives (SSDs) commonly used in desktop or laptop systems, which are available in two form factors~\cite{formfactor,nvme-express-base-specification}: Serial ATA (SATA) and M.2 Non-Volatile Memory Express (NVMe). The form factor determines the maximum achievable data throughput and the available low-level protocols for SED control.

NVMe drives use the Security Send/Receive command~\cite{nvme-express-base-specification}, while SATA drives rely on ATA command passthrough~\cite{acs-3}. These protocols serve as transport wrappers for SED commands. Unfortunately, not every driver or adapter forwards them correctly to the attached drive. The Linux kernel \emph{libata} interface turns off ATA forwarding of security commands by default, requiring an additional kernel option to enable it.
Based on our testing, USB adapters often do not forward SED commands to the drive, so the only reliable method for using external NVMe SED drives is through Thunderbolt (USB4) adapters.

\subsection{Data throughput}\label{sec:dt}
SATA performance is constrained by the underlying protocol. The most recent SATA III specification supports a theoretical maximum throughput of 6~Gb/s, which translates to a practical throughput of approximately 500~MB/s.
NVMe drives connect directly to the PCI Express (PCIe) bus, whose generation determines the performance ceiling. PCIe Gen5 supports flash storage data access at up to 12~GB/s, depending on the PCIe link speed~\cite{haas2023modern}.

Software encryption can saturate the SATA throughput, making the underlying storage hardware the performance bottleneck. In contrast, high-performance NVMe drives face a significant bottleneck in the software encryption.

\subsection{Manufacturers and firmware}\label{sec:fw}
Major manufacturers of flash-based storage include Kingston, Kioxia (formerly Toshiba), Micron, Samsung, Seagate, SK~hynix, and Western Digital (WDC). Several former manufacturers no longer operate independently: SanDisk was acquired by WDC in 2016 (and separated again in 2025), LiteOn was acquired by Kioxia in 2018, and Intel sold its storage business to SK~hynix in 2021 (now operating as Solidigm). Laptop manufacturers ship OEM versions or apply own branding. For example, Crucial products were manufactured by Micron, which retired the brand from the market in 2025. These distinctions matter when a user needs to update firmware or verify whether the drive is still being maintained. Updates for OEM drives are provided only through the support sites of the respective laptop manufacturers. If the drive was purchased separately as a second-hand unit or spare part, it is often unclear where to obtain updates.

On Linux, firmware updates should be handled through the Linux Vendor Firmware Service (LVFS)~\cite{lvfs}, which avoids the issues described above. Unfortunately, only major laptop manufacturers provide this option for OEM firmware. In other cases, the only way to update firmware is through vendor-specific Windows update tools.
Several reported vulnerabilities were never fixed, and vendors simply marked the SED capability of the drive as no longer supported, suggesting software encryption instead~\cite{lenovosde,sandisk-x300s}.

\subsection{TCG standards}\label{sec:standards}
The SED-related standards are maintained by the TCG Storage Workgroup.
SED functionality is abstracted through Security Subsystem Classes (SSCs), which extend the TCG Storage Architecture Core Specification~\cite{tcgcore}. Opal2~\cite{tcgopal} and Pyrite2~\cite{tcgpyrite}, are designed for personal use, while others target enterprise server environments. Pyrite2 provides only user authentication, whereas Opal2 also handles data encryption and specifies a broad set of features.

An Opal2 administrator can always unlock the configured LR or add access rights for other users. The Single-User Mode (SUM)~\cite{tcgsum} feature then modifies the access model so that the drive administrator can configure the LR in such a way that only the user can unlock (decrypt) the data. Administrators then can only destroy the allocated area and repurpose the allocated space.

Anyone, including ransomware, can take ownership of SEDs in the initial configuration, set up their password, and lock the drive. TCG defines the Block SID~\cite{tcgblocksid} extension that can block such ownership transition after boot, but it requires proper configuration by the user (usually in BIOS).

Vendors also frequently ship the same hardware with different firmware configurations. A~typical source of confusion is Pyrite2 versus Opal2 SSC support, where the difference may be indicated only by a subtle variation in the product label or description.
We have repeatedly encountered Pyrite2 drives marketed as providing hardware encryption. During the initial assembly of our testbed, 12~additional drives turned out to be Pyrite2 devices and were excluded from our testing.

\section{LUKS2 case study}\label{sec:luks2}

LUKS~\cite{luks1} is a format for storing metadata about disk encryption configuration. It implements keyslots, each containing a volume key for the encrypted data area. LUKS2~\cite{luks2} introduces improved keyslot protection using memory-hard algorithms and adds the concept of tokens, enabling device unlocking to be integrated with external systems, such as the Trusted Platform Modules. Once a keyslot is successfully unlocked, the volume key is used to configure a virtual device through the Linux kernel driver. LUKS2 is not designed to provide plausible deniability; its metadata are stored in plaintext and must always be readable.

Since a LUKS2 keyslot can store an arbitrary key, the keyslot key can serve as a password to unlock an SED Locking Range (LR). Given that LUKS2 keyslots are protected in the same manner as in software encryption using strong key derivation and keys are generated by a strong random number generator (RNG), we do not need to analyze the internals of SED password processing. None of the SED drives available to us publicly documents how the password is processed internally by the drive.

Our analysis investigates whether the integration of SEDs ensures that the following security goals are met:
\begin{itemize}
\item \emph{The SED behaves according to TCG specifications subset.}
\item \emph{The SED allows configuration of a LR while simultaneously providing sufficient unencrypted space for LUKS2 metadata.}
\item \emph{Once the drive is configured, all subsequent changes must be authenticated.}
\item \emph{User data inside the LR are encrypted using a suitable encryption mode.}
\item \emph{All SED administrator roles should be separated from LUKS2 unlocking. An SED administrator should not be able to bypass LUKS2 keyslot decryption.}
\item \emph{A user-friendly reset mechanism exists (destroying data and restoring the default configuration) without knowledge of any LUKS2 keyslot passwords.}
\end{itemize}

\clearpage
\subsection{Threat model}\label{sec:threat-models}
A major pitfall of disk encryption is that no widely accepted threat model exists for either software encryption or SED standards.
The absence of at least a semi-formally defined threat model is a significant shortcoming.
Several publications try to formally describe disk encryption~\cite{univis91449997,Gjosteen2005,Khati2017}, but no common model is used in practice.

Throughout this work, the following simplified threat model (security goals) applies:
The LUKS2 security goal is \emph{``The primary goal for LUKS2 is to provide data confidentiality for user data (user-friendly access to encrypted data). The secondary goal is to provide availability of the stored metadata in the case of partial and random metadata corruption.''}~\cite{luks2}.

We consider the LUKS2~\cite{luks2} security goals compatible with those of Opal2, which is designed to \emph{``protect the confidentiality of stored user data against unauthorized access once it leaves the owner's control (following a power cycle and subsequent deauthentication).''}~\cite{tcgopal}.

We assume that the attacker can access and communicate with the drive using the same low-level protocols as the user, and that the drive never returns to service after analysis by an attacker. Data is protected only when the system is powered off, and cryptographic data integrity cannot be guaranteed, as encryption preserves the plaintext length~\cite{wong2021real}.

The user must trust both the hardware vendor and the firmware, which could (e.g., via supply-chain attacks) perform covert activities, such as exfiltrating the encryption key, detecting keywords in plaintext, and storing logs in designated areas on drive for later collection. We do not cover scenarios in which the attacker can tamper with the drive or the user's system.

Initially, SEDs were promoted not only as a means to improve encryption performance but also as a superior alternative to software disk encryption, since embedding encryption in hardware-isolated systems could enable SEDs to offer stronger security properties. However, several analyses have concluded that SEDs provide no security advantage over software disk encryption~\cite{meijer2019self, sde-risks, muller2014systematic}.

To our knowledge, no formal publication has specifically addressed the security and compatibility analysis of the Opal2 drives interface itself.

Unlike software implementations, SEDs do not store encryption keys in host memory. Unfortunately, operating systems, including Linux, store the password for unlocking the SED, which is required to resume from sleep mode~\cite{meijer2019self}.
This means that an attacker who gains access to a suspended system can recover the SED password from a memory snapshot.

An attacker with physical access to an unlocked SED can disconnect and reconnect the physical data interface to the attacker's system while maintaining the drive's power supply. Host memory analysis for locating stored unlocking passwords, as well as direct memory analysis of the SED controller, is also feasible~\cite{sde-risks, muller2014systematic}.

Older drives expose debugging interfaces such as JTAG, or allow them to be re-enabled through techniques like power glitching, as storage controllers are typically not designed as security devices with tamper-detection capabilities~\cite{meijer2019self}.

\section{SED black-box analysis}\label{sec:testing}

We treat the SEDs as black boxes with documented Opal2 functionality, with no intention of probing for vendor-specific extensions.
Our functionality and security tests can be easily replicated and extended to new drives.

\subsection{SED testbed and setup}

Our testbed consists of 38 Opal2 SEDs spanning major manufacturers.
We initially collected 54 drives in total, but excluded 12 that supported only Pyrite2 and 4 with duplicate hardware revisions.
Used drives are listed in Appendix~\ref{sec:drives} and we refer to them by the identifiers (derived from vendor names) listed in Figure~\ref{tab:opal-drives}.
Several SEDs in our testbed are Original Equipment Manufacturer (OEM)  parts sourced from laptops and purchased separately.

\begin{wrapfigure}[11]{l}{0.4\textwidth}
     \raisebox{0pt}[\dimexpr\height-1.4\baselineskip\relax]
     {\includegraphics[width=0.95\linewidth]{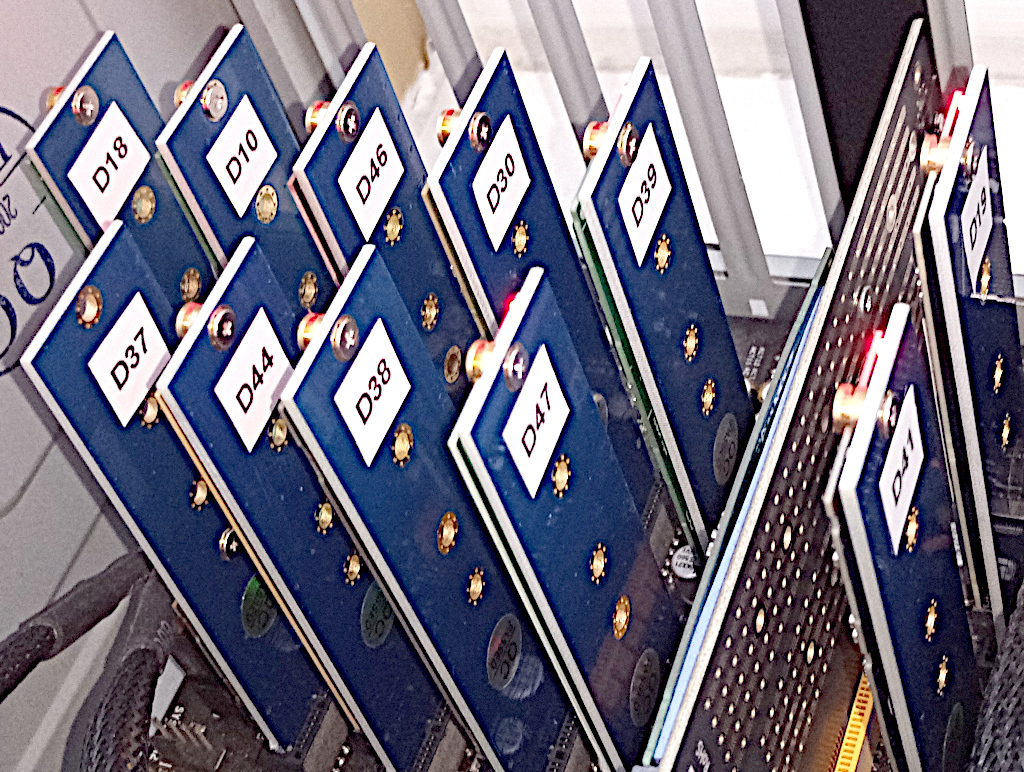}}
     \caption{NVMe testbed.}
     \label{fig:disks}
\end{wrapfigure}

All NVMe drives were connected directly to the PCIe bus via a passive adapter to eliminate any influence from intermediate controllers.
SATA drives were connected to the on-board SATA controller.
Software environment was Fedora Linux with kernel 6.19. Results should be reproducible on any operating system, as they are derived from responses returned directly by the drives. For the LUKS2 SUM extension test, we used a kernel with our patches (now merged in kernel version 7.1).

For SED configuration, we extended the \emph{Opal toolset}~\cite{toolset} and used it as the low-level layer for our \emph{Opal Test Suite}~\cite{testsuite},  which automates what was initially a manual analysis process. All tools and tests are released under free open-source licenses.
For NVMe drives configuration, \emph{nvme-cli}~\cite{nvme-cli} was used.

To avoid ethical concerns related to residual user data, all SEDs were  sanitized before testing: we performed a Physical Presence Security
Identification (PSID) reset to the factory state (see Section~\ref{sec:psid}), overwrote all data with zeroes, and issued a discard command.

\subsection{SED feature scan}\label{sec:features}

The available SSC features of an SED can be detected using the \emph{Discovery0} command~\cite{tcgcore,tcgopal}. \emph{Discovery0} is a mandatory unauthenticated command that every drive supporting any TCG SSC profile must implement; its response must contain information about all supported features. The response is parsed by the \emph{Opal toolset}~\cite{toolset}. Generic drive properties, such as support for 4096-byte sectors, are obtained using tools such as \emph{lsblk} and \emph{nvme tools}.

The features announced by our testbed drives relevant to our use case are summarized in Figure~\ref{tab:opal-features}.

\subsection{Security analysis}\label{sec:analysis}

TCG standards are extensive, but they do not provide any reference implementations, giving manufacturers considerable freedom in their implementations. Indeed, we observed that functions behave slightly differently across SEDs.
In the following analysis, we focused only on aspects relevant to the LUKS2 integration use case. This resulted in four selected tests described in detail:

\begin{itemize}
\item Configuring a LR and, if the drive does not wipe old data, checking for patterns after a key change (Section~\ref{sec:lr}).
\item Analyzing the exported output of the \emph{Random} method (Section~\ref{sec:rng}).
\item Testing the SUM feature (Section~\ref{sec:sum}).
\item Resetting the SED to the factory state using PSID (Section~\ref{sec:psid}).
\end{itemize}

\subsection{LR encryption test}\label{sec:lr}
While LR configuration is a core SED feature, advanced configurations (e.g., the offsets mandated by LUKS2) might not always work properly. Furthermore, implementation mistakes could cause encryptions and decryptions to leave undesirable patterns in the data. While Opal2 does not allow us to access the ciphertext directly, we can bypass this by analyzing the content of the unlocked area after the encryption key is changed. This is because the original ciphertext is ``decrypted'' by the new key and returned to the user.

Our LR test covers the basic LR configuration and analysis in these steps:
\begin{enumerate}
\item Configure the drive to use 512-byte sectors,
\item reset the drive with PSID to set initial conditions,
\item initialize the drive by setting user and admin passwords,
\item create a LR with a 16~MB offset with a length of 100~MB,
\item write a pattern of all ones (\emph{0xFF}),
\item lock and unlock the LR to verify written data,
\item change the LR key with TCG Core mandatory \emph{GenKey} method~\cite{tcgcore},
\item check the LR content for zeroes (i.e., no pattern detection possible),
\item run randomness analysis over the data, and if it fails,
\item perform manual inspection and comparison of patterns.
\end{enumerate}

\subsubsection{Encryption mode pattern detection}
Modern Opal2 SEDs typically use the AES-XTS encryption  mode~\cite{IEEE1619-2025}, which is designed to support parallel encryption of AES blocks after precomputing tweak values.
A critical requirement is that each encrypted sector is tweaked with a unique value to prevent ciphertext patterns from leaking information about the underlying plaintext~\cite{nistxts,xtsscopes}. The tweak is typically derived from the physical sector number, but may also be based on the internal storage architecture.

\begin{figure}[H]
\begin{floatrow}
\ffigbox[.35\textwidth]{%
    \renewcommand{\arraystretch}{0.95}
    \begin{tabular}{cccccccccccc}
    \rot{\textbf{Drive ID}} &
    \rot{SUM} &
    \rot{Block SID} &
    \rot{LR alignment} &
    \rot{4kB sector} &
    \rot{Datastore} &
    \rot{Data removal} &
    \rot{Secure msg} &
    \rot{PIN 128} \\
    \midrule
    HYN-1  & \yes & \yes & \na  & \na  & \na  & \na  & \na & \na  \\
    HYN-2  & \yes & \yes & \yes & \na  & \na  & \na  & \na & \na  \\
    HYN-3  & \yes & \yes & \na  & \na  & \yes & \na  & \na & \na  \\
    HYN-4  & \yes & \yes & \na  & \na  & \yes & \na  & \na & \na  \\
    KIN-1  & \yes & \yes & \na  & \na  & \na  & \na  & \na & \na  \\
    KIO-1  & \yes & \na  & \na  & \na  & \na  & \na  & \na & \na  \\
    KIO-2  & \yes & \yes & \na  & \yes & \na  & \na  & \na & \na  \\
    KIO-3  & \yes & \yes & \na  & \yes & \na  & \na  & \na & \na  \\
    LEN-1  & \na  & \yes & \yes & \na  & \na  & \na  & \na & \na  \\
    LEN-2  & \na  & \yes & \yes & \na  & \na  & \na  & \na & \na  \\
    LEN-3  & \yes & \yes & \na  & \na  & \na  & \na  & \na & \na  \\
    MIC-1  & \yes & \yes & \na  & \na  & \na  & \na  & \na & \na  \\
    MIC-2  & \yes & \yes & \na  & \na  & \na  & \na  & \na & \na  \\
    MIC-3  & \yes & \na  & \yes & \na  & \na  & \na  & \na & \na  \\
    MIC-4  & \yes & \yes & \yes & \yes & \yes & \yes & \na & \na  \\
    MIC-5  & \na  & \yes & \yes & \na  & \na  & \na  & \na & \na  \\
    MIC-6  & \yes & \yes & \yes & \na  & \na  & \na  & \na & \na  \\
    SAM-1  & \na  & \na  & \na  & \na  & \na  & \na  & \na & \na  \\
    SAM-2  & \yes & \na  & \na  & \na  & \na  & \na  & \na & \na  \\
    SAM-3  & \yes & \na  & \na  & \na  & \na  & \na  & \na & \na  \\
    SAM-4  & \na  & \na  & \yes & \na  & \na  & \na  & \na & \na  \\
    SAM-5  & \na  & \na  & \yes & \na  & \na  & \na  & \na & \na  \\
    SAM-6  & \na  & \na  & \yes & \na  & \na  & \na  & \na & \na  \\
    SAM-7  & \na  & \na  & \yes & \na  & \na  & \na  & \na & \na  \\
    SAM-8  & \na  & \yes & \yes & \na  & \na  & \na  & \na & \na  \\
    SAM-9  & \na  & \yes & \yes & \na  & \na  & \na  & \na & \na  \\
    SAM-10 & \na  & \yes & \yes & \na  & \na  & \na  & \na & \na  \\
    SAM-11 & \yes & \yes & \yes & \na  & \na  & \na  & \na & \na  \\
    SAM-12 & \na  & \na  & \yes & \na  & \na  & \na  & \na & \na  \\
    SAM-13 & \yes & \yes & \yes & \na  & \yes & \yes & \na & \na  \\
    SEA-1  & \yes & \yes & \yes & \yes & \yes & \yes & \na & \na  \\
    WDC-1  & \yes & \na  & \yes & \na  & \na  & \na  & \na & \na  \\
    WDC-2  & \na  & \na  & \yes & \na  & \na  & \na  & \na & \na  \\
    WDC-3  & \na  & \yes & \na  & \na  & \na  & \na  & \na & \na  \\
    WDC-4  & \yes & \yes & \yes & \yes & \na  & \na  & \na & \na  \\
    WDC-5  & \yes & \yes & \na  & \yes & \na  & \na  & \na & \na  \\
    WDC-6  & \yes & \yes & \yes & \yes & \na  & \na  & \na & \na  \\
    WDC-7  & \yes & \yes & \na  & \yes & \yes & \yes & \na & \na  \\
    \bottomrule
    \end{tabular}
}{%
     \caption{SED feature support.}
     \label{tab:opal-features}
}
\hfill
\ffigbox[.62\textwidth]{%
\parbox{.62\textwidth}{%

\xsect{SUM}
Advertised Single User Mode~\cite{tcgsum}.

\xsect{Block SID}
A mechanism used to block SED authentication. Users cannot take ownership of an unconfigured drive until the next power cycle~\cite{tcgblocksid}.

\xsect{LR alignment}
The drive requires the LR start to be aligned to a larger sector boundary (typically 4096 bytes).

\xsect{Support for 4kB sector}
The drive advertises sectors of 4096 bytes. Switching from 512-byte to 4096-byte sectors should reduce the encryption overhead.

\xsect{Datastore}
The drive allows storing application-specific metadata~\cite{tcgadditionaldatastore}.

\xsect{Supported Data Removal Mechanism}
This feature defines the supported mechanisms for erasing data~\cite{tcgopal}.

\xsect{Secure messaging}
The drive supports pre-shared key secure messaging to establish an encrypted communication channel between the SED and the operating system~\cite{tcgsecuremessaging}.

\xsect{PIN 128}
Advertised support for C\_PIN (password) enhancements including increasing maximal length from 32 to 128 characters~\cite{tcgcpin}.

\vspace{1em}
\hspace{0.8em}Figure~\ref{tab:opal-features} omits certain mandatory features, as they are always present:

\xsect{TPer feature}
(Trusted Peripheral) defines communication parameters~\cite{tcgcore},

\xsect{Locking feature}
specifies the implemented functions, including media encryption~\cite{tcgopal},

\xsect{Geometry feature}
reports the logical encryption block size and required offset alignment~\cite{tcgopal}, and

\xsect{PSID}
resets the drive to the initial factory state~\cite{tcgpsid}.

\vspace{0.4em}
Drives implement only a small subset of the optional Opal2 features. The case of secure messaging is particularly striking---none of our SEDs implements it. Another notable finding is that no drive supports longer passwords (details in Section~\ref{sec:psid}). More features are appearing in newer SEDs that claim to support Opal version 2.02.
}}{}
\end{floatrow}
\end{figure}

While all recent SEDs claim to use AES-XTS, we do not rule out the possibility that CBC (Cipher Block Chaining) or ECB (Electronic Code Book) mode~\cite{wong2021real}, or an incorrectly applied AES-XTS, is used instead.
Instances of incorrectly applied CBC or ECB modes in storage hardware have been documented previously~\cite{got-hw-crypto}.

The key observation is that encryption patterns can persist across a key change and subsequent ``decryption'' with a different key.
This test is not applicable to recent drives, as they present only zeroed content after a key change.
For drives where the test works, it reveals interesting details about the internal implementation.

The principle of our developed pattern detection is described in detail in  Appendix~\ref{sec:patterndesc} and can be generalized to cover multiple sectors.
We initially performed manual inspection of AES blocks on drives where randomness checks failed.
For randomness check the \emph{Dieharder}~\cite{dieharder} and \emph{TestU01}~\cite{testu01} was used. Our toolkit then includes a developed simplified randomness test (as described in Section~\ref{sec:rng}) to trigger a warning if a pattern is detected.

\subsubsection{Test results}
HYN-3, KIO-1, and SAM-1 fail to create an LR with our testing parameters. SAM-1 does not support all integer representation variants (atom  lengths)~\cite{tcgcore} in the LR configuration, making it impossible to configure the LR with the 64-bit integers required for our use case.
The other two SEDs implement only an insufficient subset of the TCG Opal2 interface.
These SEDs are therefore unusable in our  scenario and are likewise incompatible with LUKS2 integration.

The randomness test for an LR with a changed key failed for WDC-1, WDC-2, SAM-12, LEN-1, and LEN-2, each exhibiting a clear pattern of repeated
blocks.
These SEDs appear to use AES-XTS as expected, since no patterns characteristic of ECB or CBC mode were observed.

For LEN-1 and LEN-2, the evidence suggests AES-XTS with a 512-byte sector size is in use, but all sectors share the same tweak value.
As a result, identical sector content written at different offsets produces identical ciphertext.

For SAM-12, repeating patterns are detectable within consecutive groups of 8+8 512-byte sectors, similarly to the previous case, but only within
these two groups; subsequent groups appear to use distinct tweak values. This suggests the tweak is configured for a larger sector size than is
actually in use.

For WDC-1 and WDC-2, the pattern is considerably more complex. These SEDs appear to use 4096-byte sectors internally for encryption, where the first
2048~bytes are tweaked correctly (no patterns detected), but the remaining 2048~bytes exhibit patterns indicating that the tweaks from the first half
are reused. Additionally, these SEDs appear to operate with internal 128~kB blocks within which the tweak value always changes.

While these findings demonstrate that the vendors implemented the encryption mode incorrectly, practical exploitability is very limited.
The plaintext patterns could be detectacle even after key regeneration, meaning limited information about plaintext that should have been destroyed
remains recoverable.

\subsection{\emph{Random} Test}\label{sec:rng}
Opal2 SEDs are required to implement a random number generation method called \emph{Random}~\cite{tcgcore,tcgopal}.
The specification imposes no security-related requirements on randomness quality, leaving vendors entirely responsible for its implementation.
The Enterprise SSC documentation states that \emph{"The host is able to use the storage device's capability of generating highly random byte sequences to create keys or passwords. One way to achieve this is via the Random method."}~\cite{tcgenterprise}.

The disk encryption key (DEK) is always generated on the SED and never leaves it~\cite{tcgcore}. The host system has no access to DEKs, but has direct access to \emph{Random} outputs. While the relationship between these outputs and the DEK generator is not specified, insufficient randomness in the former raises doubts about the security of the latter.

To evaluate randomness of data produced by \emph{Random}, we utilized the right-tailed Chi-squared ($\chi^2$) goodness-of-fit test~\cite{moorestatistics}.
This test allows us to measure a fit between our observed sample, i. e., random data from a SED, and the expected uniform distribution. We performed following steps for each SED:
\begin{enumerate}
\item Extract a 1~MB data sample from the SED using \emph{Opal toolset}~\cite{toolset},
\item compute a histogram of individual byte values present in the extracted data sample (\emph{0x00} to \emph{0xFF} representing 256 individual categories),
\item compute a $\chi^2$ statistic using observed byte frequencies from the histogram, and
\item perform a $\chi^2$ goodness-of-fit test with 255 degrees of freedom and a significance level of 0.1.
\end{enumerate}

\begin{figure}[H]
  \centering

  \begin{minipage}[t]{0.49\textwidth}
    \centering
    \includegraphics[width=\linewidth]{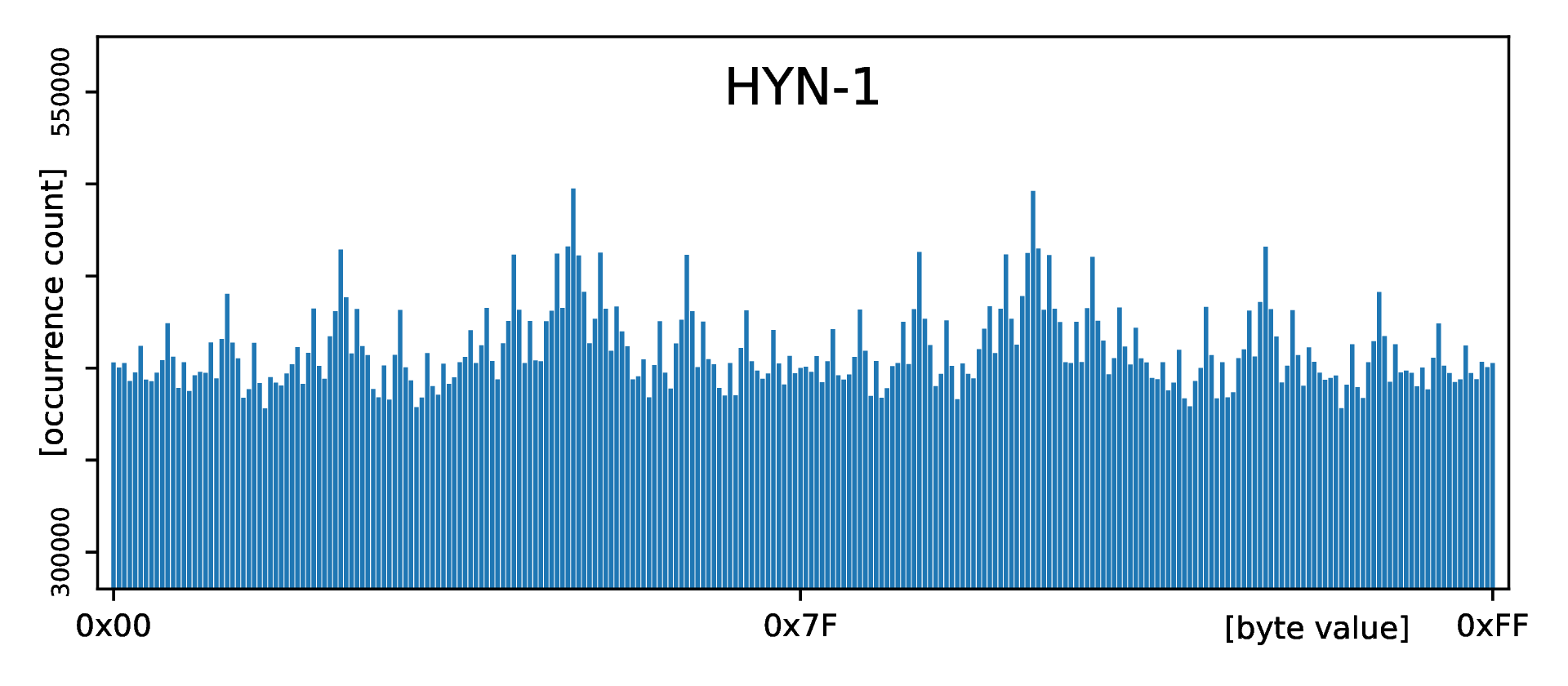}
  \end{minipage}
  \hfill
  \begin{minipage}[t]{0.49\textwidth}
    \centering
    \includegraphics[width=\linewidth]{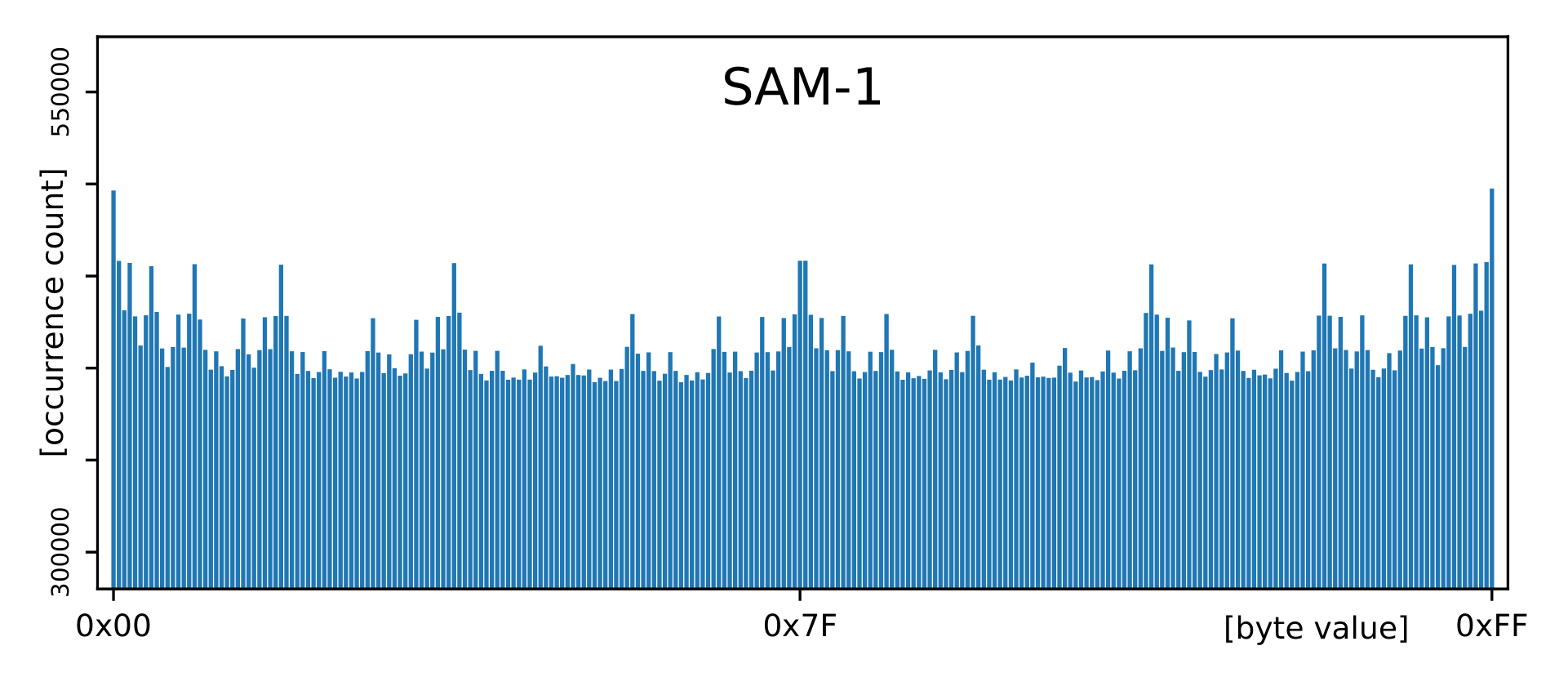}
  \end{minipage}

  \vspace{1em} 

  \begin{minipage}[t]{0.49\textwidth}
    \centering
    \includegraphics[width=\linewidth]{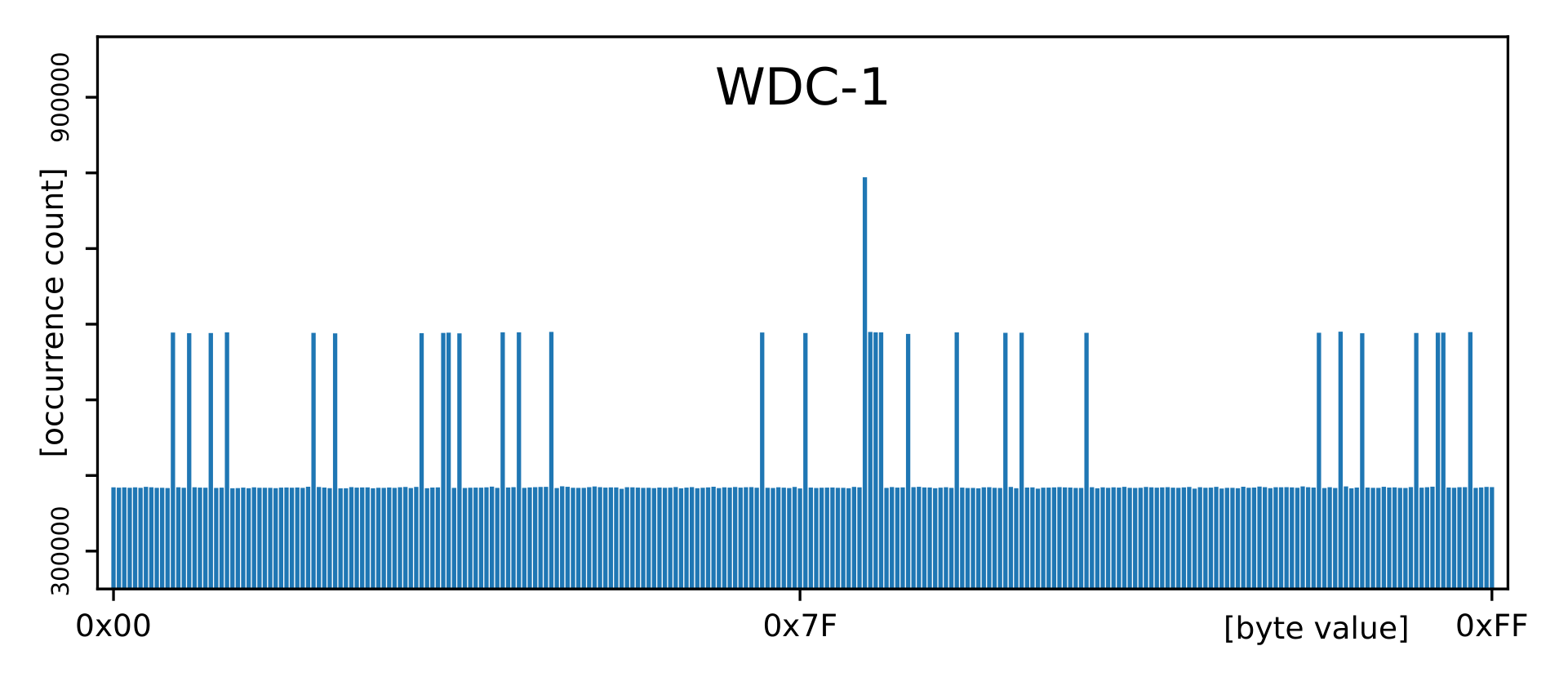}
  \end{minipage}
  \hfill
  \begin{minipage}[t]{0.49\textwidth}
    \centering
    \includegraphics[width=\linewidth]{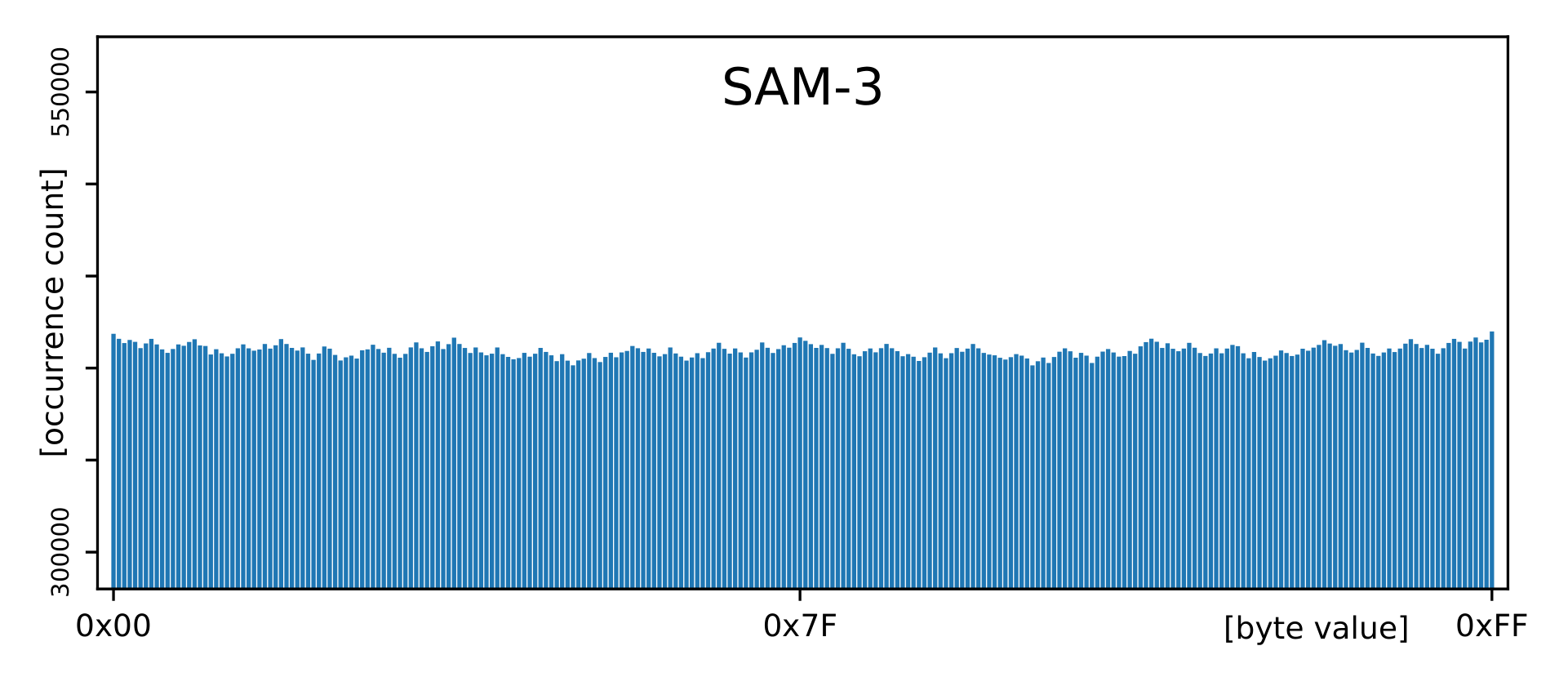}
  \end{minipage}

  \caption{Histograms of byte values of \emph{Random} outputs with a detectable bias.}
    \label{fig:histograms}
\end{figure}

Random data extraction speeds varied considerably across SEDs, ranging from hundreds of bytes to hundreds of kilobytes per second, reflecting the diversity of RNG implementations across vendors.

\subsubsection{Test results}
HYN-1, LEN-1, LEN-2, SAM-1, SAM-2, SAM-3 and WDC-1 fail the randomness tests, producing statistically non-uniform outputs. Most notably, LEN-1 and LEN-2 generate highly predictable outputs -- a recurring sequence of values from 0 to 32 for each \emph{Random} call.

We examined the failing samples in greater detail by inspecting their byte histograms, which record the number of occurrences of each possible byte value. For a well-functioning randomness source, this distribution should be uniform. The histograms for HYN-1, SAM-1, SAM-3, and WDC-1 are shown in Figure~\ref{fig:histograms} and illustrate a significant lack of uniformity.

WDC-1 is strongly biased towards a small set of specific values; byte value \emph{0x8B} appears in its RNG output approximately twice as frequently as the majority of other values.

HYN-1, SAM-1, and SAM-3 also deviate substantially from uniformity, all exhibiting a symmetric distribution centred around byte value \emph{0x7F}
(i.e., the midpoint of the byte value axis). While SAM-3 shows a marked improvement over its predecessor SAM-1, both exhibit elevated frequencies
at \emph{0x00} and \emph{0xFF}. Furthermore, the histograms of SAM-1 and SAM-3 share elevated counts at the same byte values, suggesting that the
vendor identified the issue and attempted to correct the RNG implementation in SAM-3, but the underlying fix remained inadequate.

These randomness deficiencies have not appeared in any public security reports, despite prior reverse-engineering work on SAM-1~\cite{meijer2019self}.
If any user relies on the \emph{Random} method for generating cryptographic keys, it could be a critical security issue. We have not found such a real application.

Based on the results, we will not use the SED \emph{Random} method for LUKS2 either. We stick with the kernel random generator, which is already suitable for long-term cryptographic keys.

\subsection{SUM Test}\label{sec:sum}
As described in Section~\ref{sec:standards}, the SUM feature~\cite{tcgsum} implements a different threat model in which the user can have exclusive
control over LR unlocking. It modifies the Opal2 \emph{Activate} method~\cite{tcgopal} to allow SUM configuration, and introduces a new \emph{Reactivate} method~\cite{tcgsum} to enable or disable SUM for an existing LR.

We found no SUM support in any open-source tool, indicating that the feature has never been exercised by existing tooling. We added support to the \emph{Opal toolset}~\cite{toolset} and implemented a basic test that creates an LR in SUM mode and attempts to toggle SUM access on an existing LR.

The test performs the following steps:
\begin{enumerate}
\item Reset the drive via PSID to establish initial conditions,
\item initialize the drive by setting the user password for SUM,
\item create an LR with a 16~MB offset and a length of 100~MB in SUM mode,
\item lock and unlock the LR as the configured user,
\item verify that the admin user cannot unlock the LR,
\item disable SUM for the LR using the \emph{Reactivate} method,
\item attempt to unlock the LR as the admin user, and
\item re-enable SUM on the LR.
\end{enumerate}

\subsubsection{Test results}
KIO-1, KIO-2, MIC-1, and MIC-2 did not pass the test, as their implementation of the \emph{Reactivate} method appears to deviate from the behavior our scripts expect based on the TCG specification~\cite{tcgsum}.

A notable property of the SUM requirements~\cite{tcgsum} concerns how they define the transition of a LR between SUM and non-SUM modes while preserving the encrypted data. The transition into SUM mode requires locking to be disabled, which effectively removes the protection of the LR. Conversely, the transition out of SUM mode resets the user password to its default value (an empty string). In both cases, an attacker capable of interrupting the process at this stage could gain access to the data without any password protection. For this reason, we have deliberately chosen not to support such transitions in the LUKS2 implementation, as they cannot be realized in a secure manner.

\subsection{PSID Test}\label{sec:psid}
An Opal2 SED ships with default configuration and requires an initial setup to establish administration credentials and define LR configuration and access rights. If those credentials are subsequently lost, whether because the user forgets the password or the SED is transferred to another user in a locked state, the only way to revert the SED to its factory state is via the Physical Presence Security Identification (PSID).

\begin{wrapfigure}[9]{l}{0.3\textwidth}
     \raisebox{0pt}[\dimexpr\height-1.4\baselineskip\relax]{\includegraphics[width=0.95\linewidth]{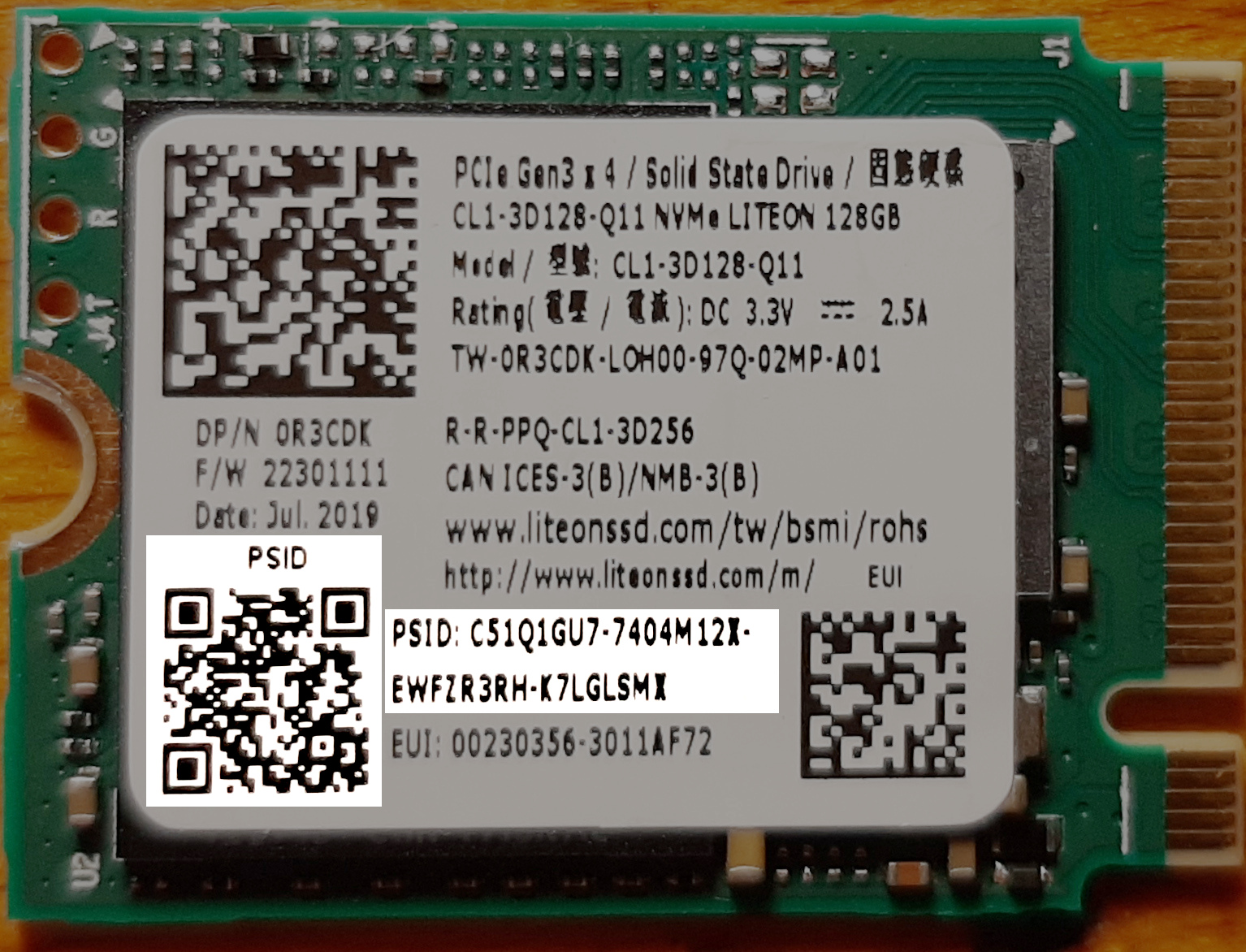}}
     \caption{PSID label.}
     \label{fig:nvme}
\end{wrapfigure}

The PSID should be obtainable only by physically accessing the SED label and must be randomly generated~\cite{tcgpsid}. In particular, it must not
be possible to derive the PSID from the serial number or any other publicly observable attribute.
Although the TCG document~\cite{tcgpsid} defines no formal requirements for PSID format or placement, all SEDs in our testbed use a consistent format of 32 alphanumeric characters unique to each drive, with the PSID printed on the label, either directly or as a QR code, as shown in Figure~\ref{fig:nvme}.

The 32-character length is in fact the maximum password length across all TCG SSCs, imposed by the underlying data type~\cite{tcgcore}. This constraint is not clearly articulated in the standards. TCG attempted to address this limitation by introducing an optional password length extension in the C\_PIN Enhancements feature~\cite{tcgcpin}, but as shown in Figure~\ref{tab:opal-features}, no drive in our testbed implements it.

A further finding concerns user awareness: users do not realise that the PSID should be kept confidential. Numerous publicly available images of SEDs with clearly visible PSIDs exist online (easily verified via a simple ``NVMe PSID'' image search). Anyone in possession of such an image and with remote access to a \emph{locked} drive could destroy its data using the exposed PSID.
This is not a serious security issue in itself, since physical access to the drive likely affords other means of data destruction. A more consequential scenario arises when the drive is embedded in a sealed laptop or its label has been destroyed: in such cases, the user may be locked out of resetting own drive without voiding the warranty.

\begin{wraptable}[41]{l}{0pt}
    \renewcommand{\arraystretch}{0.98}
    \begin{tabular}{cccccc}
    \rot{\small\textbf{{Drive ID}}} &
    \rot{\tiny\makecell[l]{\textbf{XTS pattern}\\\yes random\\\no non-random\\\,0~zeroed\\\fail LR fail}} \vrule &
    \rot{\tiny\makecell[l]{\yes RNG ok\\\no RNG fail}} \vrule &
    \rot{\tiny\makecell[l]{\yes PSID ok\\\no suffix ignored}} \vrule &
    \rot{\tiny\makecell[l]{\yes SUM ok\\\no SUM fail\\\,\na~SUM n/a}} \vrule &
    \rot{\tiny\makecell[l]{\na LUKS2 n/a \\\yes LUKS2 ok\\\yes\yes LUKS2+SUM}} \\
    \midrule
    HYN-1  & \yes  & \no  & \yes & \yes & ~~ \yes\yes \\
    HYN-2  & 0     & \yes & \yes & \yes & ~~ \yes\yes \\
    HYN-3  & \fail & \yes & \yes & \yes & ~~ \yes     \\
    HYN-4  & 0     & \yes & \yes & \yes & ~~ \yes\yes \\
    KIN-1  & 0     & \yes & \no  & \yes & ~~ \yes\yes \\
    KIO-1  & \fail & \yes & \yes & \no  & ~~ \na      \\
    KIO-2  & \yes  & \yes & \yes & \no  & ~~ \yes     \\
    KIO-3  & 0     & \yes & \yes & \yes & ~~ \yes\yes \\
    LEN-1  & \no   & \no  & \yes & \na  & ~~ \yes     \\
    LEN-2  & \no   & \no  & \yes & \na  & ~~ \yes     \\
    LEN-3  & \yes  & \yes & \no  & \yes & ~~ \yes     \\
    MIC-1  & \yes  & \yes & \no  & \no  & ~~ \yes     \\
    MIC-2  & \yes  & \yes & \no  & \no  & ~~ \yes     \\
    MIC-3  & \yes  & \yes & \no  & \yes & ~~ \yes     \\
    MIC-4  & 0     & \yes & \yes & \yes & ~~ \yes\yes \\
    MIC-5  & 0     & \yes & \no  & \na  & ~~ \yes     \\
    MIC-6  & 0     & \yes & \yes & \yes & ~~ \yes\yes \\
    SAM-1  & \fail & \no  & \yes & \na  & ~~ \na      \\
    SAM-2  & \yes  & \yes & \yes & \yes & ~~ \yes     \\
    SAM-3  & \yes  & \no  & \yes & \yes & ~~ \na      \\
    SAM-4  & \yes  & \yes & \yes & \na  & ~~ \yes     \\
    SAM-5  & \yes  & \yes & \yes & \na  & ~~ \yes     \\
    SAM-6  & \yes  & \yes & \yes & \na  & ~~ \yes     \\
    SAM-7  & \yes  & \yes & \yes & \na  & ~~ \yes     \\
    SAM-8  & \yes  & \yes & \yes & \na  & ~~ \yes     \\
    SAM-9  & \yes  & \yes & \yes & \na  & ~~ \yes     \\
    SAM-10 & 0     & \yes & \yes & \na  & ~~ \yes     \\
    SAM-11 & \yes  & \yes & \yes & \yes & ~~ \yes\yes \\
    SAM-12 & \no   & \yes & \yes & \na  & ~~ \yes     \\
    SAM-13 & 0     & \yes & \yes & \yes & ~~ \yes\yes \\
    SEA-1  & 0     & \yes & \yes & \yes & ~~ \yes\yes \\
    WDC-1  & \no   & \no  & \yes & \yes & ~~ \yes     \\
    WDC-2  & \no   & \yes & \yes & \na  & ~~ \yes     \\
    WDC-3  & 0     & \yes & \yes & \na  & ~~ \yes     \\
    WDC-4  & 0     & \yes & \yes & \yes & ~~ \yes\yes \\
    WDC-5  & 0     & \yes & \yes & \yes & ~~ \yes\yes \\
    WDC-6  & 0     & \yes & \yes & \yes & ~~ \yes\yes \\
    WDC-7  & 0     & \yes & \yes & \yes & ~~ \yes\yes \\
    \bottomrule
    \end{tabular}
    \caption{SED security evaluation.}\label{tab:opal-issues}
\end{wraptable}

To validate that PSID works as expected, we performed the following steps:
\begin{enumerate}
\item Issue PSID revert to establish initial conditions, and
\item append a character to the PSID as a suffix and repeat the command.
\end{enumerate}

\subsubsection{Test results}
All drivers support PSID, which resets the drive state to default settings.
WDC-1 had PSIDs with a common prefix and suffix generated sequentially, as verified by examining multiple SEDs from the same production batch (we had seen multiple drives).
Recovering the PSID of another drive in the batch is therefore trivially achievable by trying candidate values incrementally. This issue is absent
in WDC-2, a newer production batch of the same model, which generates PSIDs that appear random.

KIN-1, LEN-3, MIC-1, MIC-2, MIC-3, and MIC-5 implement the 32-character length check incorrectly and accept a PSID with any arbitrary suffix as valid. While this does not constitute a direct security vulnerability, it reflects poor security engineering practices.

\subsection{Other issues and test summary}
MIC-4 supports sectors of sizes both 512 and 4096 bytes. However, a sector size change did not properly propagate to the Opal2 layer, causing a mismatch between the physical and the Opal2 logical block size, even after a PSID reset. This could lead to a wrong calculation of LR units, resulting either in a loss of data, or having only a part of the LR encrypted.

SAM-11 does not allow credentials to be changed after the initial user configuration, requiring a PSID reset.

The aggregated list of findings is in Table~\ref{tab:opal-issues}. The SUM column indicates tests performed with the {Opal toolset}, while the last column indicates usability for the LUKS2 use case. The LUKS2 test is more complex, integrated into \emph{cryptsetup}. Of the 38 drives announcing Opal2 support, 3 are unusable due to firmware issues. In LUKS2 SUM mode, 24 drives report Opal2 SUM support, but only 14 can be used for LUKS2 due to firmware incompatibilities.

\subsection{Vendor notification}
We tried to responsibly disclose the relevant security issues to vendors. For WDC-1 and WDC-2, and also for Lenovo OEM drives SAM-12, LEN-1, LEN-2, and HYN-1, the issues were either already reported~\cite{sandisk-x300s,lenovosde} or we were informed that these SEDs are no longer supported.
For the sector issue in MIC-4, Micron fixed it in the newly released firmware~\cite{crucial-t500}. Reports for PSID suffix issue remain unanswered.

\section{Practical Linux applications}\label{sec:practical}
Our Opal2 SED evaluation was conducted with the goal of improving Linux open-source disk encryption to make SED drives straightforward to deploy.
As the practical output of this work, we implemented following extensions and tools:

\begin{itemize}
\item \emph{cryptsetup}~\cite{cryptsetup} Opal2 extension for LUKS2 (version 2.7), with an extension for SUM (staged for version 2.9),
\item required extensions to the Linux kernel SED interface to support SUM (staged for Linux kernel version 7.1),
\item extensions to the \emph{Opal toolset}~\cite{toolset}, including feature discovery and complete low-level handling of Opal2 drives including SUM feature, and
\item the \emph{Opal Test Suite}~\cite{testsuite}, a new automated framework to replicate the tests described in this paper.
\end{itemize}

\section{Conclusions}\label{sec:conclusion}

Disk encryption is a widely used method to protect data, and hardware-based storage encryption has been studied extensively.
While Opal2 is offered by several vendors, no prior work has systematically compared its security properties and compatibility across vendor implementations. Although our study is LUKS2 case-based and intentionally focused on specific parts of the interface, it provides real-world
evidence that current hardware implementations of encryption are still far from ideal.

We investigated 38 Opal2 SEDs to assess their compatibility with LUKS2, uncovering security flaws and numerous incompatibilities in the process.
We provide the \emph{Opal toolset}~\cite{toolset} and the \emph{Opal Test Suite}~\cite{testsuite} to automate this analysis and surface new issues in black-box devices going forward.
The major impact of this work is the integration of Opal2 support, including SUM mode, into standard Linux LUKS2 disk encryption.

While admin-isolated SUM mode represents a stronger security posture, not all drives support it.
From 38 drives in total, 24 announce SUM support, but only 14 are usable in practice for LUKS2 due to firmware incompatibilities.
We hope vendors could adopt our \emph{Opal Test Suite} to identify such issues early.

Based on these findings, the SUM feature in LUKS2, implemented in \emph{cryptsetup}~\cite{cryptsetup}, now incorporates checks to detect correct firmware implementation, with automatic fallback to non-SUM mode when necessary.
In other words, the system will automatically select the best available configuration.
These checks can be refined further as user feedback accumulates.

The \emph{Opal Test Suite} should be extended to cover more recent drives.
Assembling a representative testbed is costly, and without community involvement, and ideally vendor cooperation, comprehensive coverage is nearly impossible. We hope that publishing this research raises awareness of the value of open-source-driven hardware security evaluation.

More broadly, TCG storage standards are complex, and as our feature scan in Section~\ref{sec:analysis} shows, vendors typically implement only the bare mandatory minimum.
Not a single tested drive supported optional secure messaging, and the C\_PIN extension feature was equally absent.
A more holistic design approach to standards defining storage encryption would reduce the attack surface and limit potential security exposure.
A striking illustration is that certain Opal2 implementations were found to be entirely broken and subsequently abandoned by their vendors, who now recommend software encryption instead~\cite{sandisk-x300s,lenovosde}.

We reported all security-relevant issues to vendors through a responsible disclosure process.
Unsurprisingly, reports concerning end-of-support drives will remain unaddressed, and vendors did not attempt to reproduce or confirm the issues at all.
We must therefore assume that drives carrying problematic firmware will remain in active use.
Even when firmware updates are available, the upgrade process is often cumbersome, requiring dedicated Windows tools or special bootable images.
Wider adoption of LVFS~\cite{lvfs} would substantially improve this situation, as users would be notified of firmware updates through their regular system notifications.

Nevertheless, applications with high security requirements can already benefit today from combining SEDs with software disk encryption as a complementary layer of protection.

\section*{Acknowledgements}
Milan Brož was supported by the European Union under Grant Agreement No.~101087529 (CHESS -- Cyber-security Excellence Hub in Estonia and South Moravia), and Tamara Čierniková by the Ai-SecTools (VJ02010010) project.

We would like to thank Štěpán Horáček for the initial implementation of the \emph{Opal toolset} and anonymous reviewers and our colleagues for helpful feedback.

\clearpage
\bibliographystyle{splncs04}
\bibliography{sed-usage}

\clearpage
\appendix

\section{Pattern detection in sector encryption}\label{sec:patterndesc}
Divide the LR into sectors $SX$ for $X\in \{1,2\}$, with each $SX$ consisting of 32 AES 16-byte blocks $PX_0,\dots,\ ..\ PX_{31}$.

We fill each $PX_i$ with bytes \emph{0xFF}, except for $P2_1$, where we put different data, e.g., \emph{0x00}. Using the XTS sector tweaks $T_{SX}$, the SED encrypts $SX$ to $CX$, divided into AES blocks $CX_0, \dots, CX_{31}$.

After the LR key change, the SED will return ``decrypted'' sectors $DX$, divided into AES blocks $DX_0, \dots, DX_{31}$ (instead of the original plaintext).

{\setlength{\intextsep}{1pt}
\begin{figure}[ht]
    \includegraphics[width=0.9\linewidth]{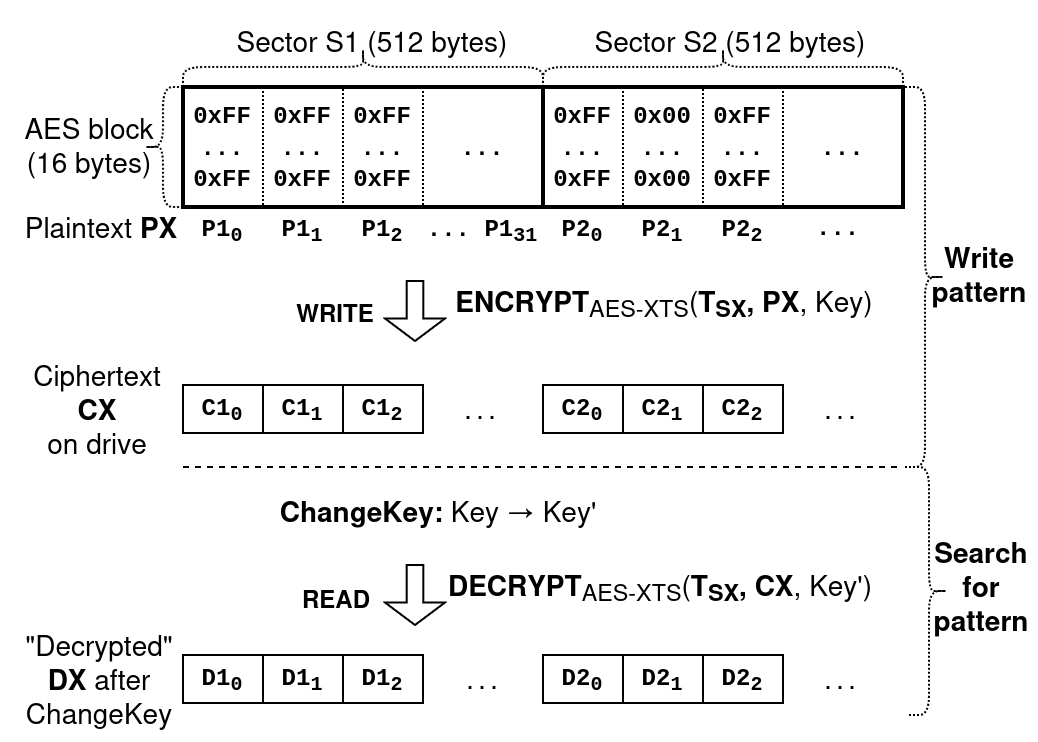}
    \centering
    \caption{The principle of detecting sector encryption patterns with AES-XTS mode. The structure of the picture is still the same if CBC or ECB is used (though the first argument $T_{SX}$ is either replaced with an $IV_{SX}$, or dropped, respectively).}
    \label{fig:pattern}
\end{figure}
}

Now let us consider several cases, depending on the actual encryption mode being used by the implementation:
\begin{itemize}
\item XTS with unique tweak values $T_{SX}$ for each sector, or CBC with unique $IV_{SX}$ for each sector. Then all $DX$ should contain pseudorandom data without any detectable patterns (and we cannot determine which mode is used).
\item XTS with $T_{S1} = T_{S2}$. Then $D1_i = D2_i$ for all $i \neq 1$, while $D1_0 \neq D1_1$ (because already $C1_i = C2_i$ for all $i \neq 1$, while $C1_0 \neq C1_1$, as the handling of each block depends on its position within the sector, but not on other blocks, for both encryption and decryption).
\item CBC with $IV_{S1} = IV_{S2}$. Then $D1_0 = D2_0$, while $D1_i = D2_i$ for all $i \geq 1$ (because already $C1_0 = C2_0$, while $C1_i = C2_i$ for all $i \geq 1$, as each block depends on the previous one for both encryption and decryption).
\item ECB. Then $D1_i = D2_j$ for all $i$ and all $j\neq 1$ (because already $C1_i = C2_j$ for all $i$ and all $j\neq 1$, as all blocks are handled independently of their position for both encryption and decryption).
\end{itemize}

\clearpage
\section{Tested drives}\label{sec:drives}
{\setlength{\intextsep}{1pt}
\begin{table}[H]
\renewcommand{\arraystretch}{1.1}
\begin{tabular}{cccccc}
\textbf{ID} & \textbf{Vendor and model} & \textbf{Type} & \textbf{Firmware} & \textbf{GB} \\
\midrule
HYN-1  & Intel SSDPEKKF256G7L                & NVMe     & 121P     &  256 \\
HYN-2  & Intel SSDPEKKF256G8L                & NVMe     & L08P     &  256 \\
HYN-3  & SK Hynix HFS256GDE9X081N            & NVMe     & 41720C20 &  256 \\
HYN-4  & SK Hynix HFS001TEJ9X162N            & NVMe     & 51730A10 & 1024 \\
KIN-1  & Kingston SKC600512G                 & SATA     & S4800105 &  512 \\
KIO-1  & Toshiba THNSFJ256GCSU               & SATA     & JULA1102 &  256 \\
KIO-2  & Kioxia KBG4AZNS256G                 & NVMe     & AEGA1102 &  256 \\
KIO-3  & Kioxia KXG6AZNV256G                 & NVMe     & 5107AGLA &  256 \\
LEN-1  & Lenovo LENSE20256GMSP34MEAT2TA      & NVMe     & 2.6.8341 &  256 \\
LEN-2  & U.Memory LENSE30256GMSP34MEAT3TA    & NVMe     & 2.5.0412 &  256 \\
LEN-3  & UnionMemory RPETJ256MGE2MDQ         & NVMe     & 1.3Q0630 &  256 \\
MIC-1  & Crucial CT250MX500SSD1              & SATA     & M3CR020  &  250 \\
MIC-2  & Crucial CT250MX500SSD4              & M.2 SATA & M3CR023  &  250 \\
MIC-3  & Micron MTFDDAK256MAY                & SATA     & M5T4     &  256 \\
MIC-4  & Crucial T500 CT1000T500SSD8         & NVMe     & P8CR004  & 1024 \\
MIC-5  & Micron 2400 MTFDKBA512QFM           & NVMe     & V3MA002  &  512 \\
MIC-6  & Micron 2450 MTFDKBA512TFK           & NVMe     & V5MA010  &  512 \\
SAM-1  & Samsung SSD 840 EVO                 & SATA     & EXT0DB6Q &  250 \\
SAM-2  & Samsung SSD 850 EVO                 & SATA     & EMT02B6Q &  500 \\
SAM-3  & Samsung SSD 850 PRO                 & SATA     & EXM04B6Q &  512 \\
SAM-4  & Samsung SSD 860 EVO                 & M.2 SATA & RVT24B6Q &  500 \\
SAM-5  & Samsung SSD 860 EVO                 & SATA     & RVT03B6Q & 1024 \\
SAM-6  & Samsung SSD 870 EVO                 & SATA     & SVT02B6Q &  500 \\
SAM-7  & Samsung SSD 960 EVO                 & NVMe     & 2B7QCXE7 &  250 \\
SAM-8  & Samsung SSD 970 EVO Plus            & NVMe     & 2B2QEXM7 &  500 \\
SAM-9  & Samsung SSD 980                     & NVMe     & 2B4QFXO7 &  500 \\
SAM-10 & Samsung SSD 980 PRO                 & NVMe     & 5B2QGXA7 & 2048 \\
SAM-11 & Samsung SSD 990 PRO                 & NVMe     & 4B2QJXD7 & 1024 \\
SAM-12 & Samsung MZVKW512HMJP-000L7          & NVMe     & 6L6QCXA7 &  512 \\
SAM-13 & Samsung SSD 9100 PRO                & NVMe     & 0B2QNXH7 & 8192 \\
SEA-1  & Seagate FireCuda 540 ZP1000GM30004  & NVMe     & SUESR101 & 1024 \\
WDC-1  & SanDisk SD7UB2Q512G1122             & SATA     & X2180300 &  512 \\
WDC-2  & SanDisk SD7TB3Q-128G-1006           & SATA     & X2180306 &  128 \\
WDC-3  & SanDisk SD9TB8W-512G-1006           & SATA     & X6104106 &  512 \\
WDC-4  & WDC PC SN730 SDBQNTY-256G-1001      & NVMe     & 11170101 &  256 \\
WDC-5  & WDC PC SN740 SDDQMQD-256G-1001      & NVMe     & 73046101 &  256 \\
WDC-6  & WDC PC SN740 SDDQNQD-512G-1014      & NVMe     & 73101100 &  512 \\
WDC-7  & WDC PC SN5000S SDEQNSJ-2T00-1002    & NVMe     & 34230100 & 2048 \\
\bottomrule
\end{tabular}
\caption{Tested Opal2 drives.}\label{tab:opal-drives}
\end{table}
}
\end{document}